\documentclass[12pt,a4paper,twoside]{article}

\usepackage{amsmath}
\usepackage{amssymb}

\numberwithin{equation}{section}
\newtheorem{theorem}{Theorem}[section]

\newtheorem{example}[theorem]{Example}

\newenvironment{proof*}{\paragraph{Proof.}}{}

\newcommand{\hk}{\hslash}
\newcommand{\arrow}{\rightarrow}
\newcommand{\map}{\mapsto}
\newcommand{\arr}{\longleftrightarrow}
\newcommand{\bb}[1]{\mathbb{#1}}
\newcommand{\Diff}[3]{\left . \frac{d}{d#2}#1\right |_{#3}}

\newcommand{\alg}{\mathfrak{g}}

\newcommand{\mg}{\mathfrak{a}}

\newcommand{\lo}{\mathcal{L}}
\newcommand{\Z}{\mathbb{Z}}

\newcommand{\pr}{\partial}
\newcommand{\me}{\geqslant}

\newcommand{\bra}[1]{\left (#1\right )}
\newcommand{\brac}[1]{\left [#1\right ]}
\newcommand{\pobr}[1]{\left \{#1\right \}}

\newcommand{\pd}[2]{\frac{\partial #1}{\partial #2}}

\newcommand{\eqreff}[2]{(\ref{#1}-\ref{#2})}

\newcommand{\e}{\Lambda}
\newcommand{\ti}{\textbf{\textsl{t}}}
\newcommand{\s}{\mathbb{S}}
\newcommand{\la}{\lambda}

\begin{document}

\title{Dispersionful analogue of the Whitham hierarchy}

\author{B\l a\.zej M.
Szablikowski$^\dag$ and Maciej B\l aszak$^\ddag$\\[3mm]
\small Institute of Physics, A. Mickiewicz University\\
\small Umultowska 85, 61-614 Pozna\'n, Poland\\
\small E-mails: $^\dag$bszablik@amu.edu.pl and $^\ddag$blaszakm@amu.edu.pl}
\date{}
%\eads{\mailto{blaszakm@main.amu.edu.pl} and \mailto{bszablik@amu.edu.pl}}

%\submitto{\JPA}

\maketitle

\begin{abstract}
The dispersionful analogue, by means of Lax formalism, of the
zero-genus universal Whitham hierarchy together with its algebraic
orbit finite-field reductions is considered. The theory is
illustrated by several significant examples.
\end{abstract}

\section{Introduction}

In \cite{K1,K2} Krichever introduced the so-called {\it universal
Whitham hierarchies} by means of moduli spaces of Riemann surfaces
of all genera. We are interested in {\it the zero-genus case}, which
we will henceforth refer to as the Whitham hierarchy. This hierarchy
contains such particular nonlinear integrable systems like
dispersionless limits of KdV, Toda, AKNS (or equivalently nonlinear
Schr\"odinger) soliton equations. However, the class of integrable
systems is much wider. There are very important reductions to the
Whitham hierarchy, related to the so-called algebraic orbits
\cite{K2,GMMA}, leading to the construction of (1+1)-dimensional
integrable dispersionless systems that are generated by rational Lax
functions. Thus, there arises a question of the construction of a
theory analogous to the Whitham hierarchy leading to the soliton
systems, i.e. systems with dispersion. The main reason for the lack of
such a complete theory are problems with the notion of higher order
"finite" poles in the case of algebra of pseudo-differential
operators.

The part of the Whitham hierarchy related to the puncture at
infinity is nothing but dispersionless limit of the well-known
infinite-field KP hierarchy. In \cite{K3} Krichever considered
reductions of KP hierarchy generated by rational
(pseudo-differential) Lax operators being quotients of appropriate
purely-differential operators. Besides, he has shown that there
exist additional symmetries that are not themselves reductions of
KP hierarchy. From the more general point of view 'quantization' of
dispersionless systems from the Whitham hierarchy generated by
rational Lax functions with "finite" poles of first order is
solved in \cite{EOR}. This solution is given by means of Lax
operators of KdV type with the additional so-called source terms
\cite{OS}. This class of integrable soliton systems contains,
among others, full AKNS hierarchy. In the series of articles
\cite{BX}-\cite{BX2} Bonora and Xiong investigated integrable
differential-difference hierarchies of the Toda type arising in
the context of the multi-matrix models. They showed how this
hierarchies, using one flow, can be turned into completely
equivalent purely differential hierarchies that can be recognized
as special rational reductions of the KP hierarchy. They also considered
dispersionless limit of the above integrable systems.

Let us illustrate such a procedure on the simplest example. The Toda
hierarchy, being a hierarchy of mutually commuting lattice soliton
equations, has a Lax formulation expressed by shift operators. The
pair of linear Lax equations have the form
\begin{align}
\label{t1} &\Psi_{n+1} + u_n\Psi_n + v_n\Psi_{n-1} = \lambda \Psi\\
\label{t2} &\pr_x \Psi_n = \Psi_{n+1} + u_n\Psi_n,
\end{align}
where the second equation defines the evolution of the eigenfunction
$\Psi$. The compatibility between equations \eqref{t1} and
\eqref{t2} leads to the Toda equation. It is known that the above
auxiliary linear problem can be rewritten with the use of
pseudo-differential operators as it follows from \eqref{t2} that
$\Psi_{n+1} = (\pr_x + u_n)\Psi_n$ and $\Psi_{n-1} =  (\pr_x +
u_{n-1})^{-1}\Psi_n$ \cite{BX},\cite{UT}-\cite{CDZ}. Then, the
linear equation \eqref{t1} takes the form
\begin{equation*}
\bra{\pr_x  + v_n(\pr_x-u_{n-1})^{-1}}\Psi = \lambda \Psi
\end{equation*}
of the eigenvalue problem related to the AKNS system. Consequently,
the full Toda hierarchy in the representation of pseudo-differential
operators takes the form of the well known AKNS hierarchy.
Following the above idea in the opposite direction one
is able to construct, with the aid of shift operators, a dispersionful
analogue of the Whitham hierarchy.

Very recently Takasaki considered dispersionless limit of
multi-component KP hierarchy with charges (lattice variables),
that can be treated as a generalization of the Toda hierarchy \cite{T}. In this
sense the original Toda hierarchy is equivalent to the
two-component KP hierarchy. Rewriting the multi-component KP
hierarchy in the scalar Lax formalism Takasaki showed that the
above dispersionless limit is the Whitham hierarchy \cite{T}. Further in
\cite{TT1} Takasaki and Takebe analysed the above issue from the
point of view of Hirota equations and Fay identities.

The theory of Whitham hierarchy and its dispersionful analogue is
inseparably connected with problems of systematic constructions
of integrable nonlinear dynamical systems. It is well known that a
very powerful tool, called the classical $R$-matrix formalism, can
be used for systematic construction of (1+1)-dimensional field and
lattice integrable dispersive systems (soliton systems)
\cite{STS}-\cite{BS} as well as dispersionless integrable field
systems \cite{TT}-\cite{SB}. Moreover, the $R$-matrix approach
allows a construction of Hamiltonian structures and conserved
quantities.

The aim of this article is to consider the dispersionful
analogue of the Whitham hierarchy and its algebraic orbit reductions
on the level of the explicit Lax representations. The article is divided
onto two parallel parts concerning dispersionless and dispersive cases, respectively. First the basic facts about the Whitham hierarchy
together with its algebraic orbits, given by some meromorphic functions, are presented. Next an alternative formulation of the original Whitham hierarchy is constructed. This formulation is the one that in the formal quantization procedure
yields the dispersionful Whitham hierarchy. Initially the dispersionful Whitham
hierarchy is presented by means of pseudo-differential operators, but the one-to-one analogue of the original Whitham hierarchy is given with the simultanous use of the shift operators following reference \cite{T}.
This is a multi-component Toda-like formulation of the dispersionful Whitham hierarchy. The finite-field reductions being counterparts of algebraic orbits are also considered. The whole theory is illustrated by the rich set of significant examples. Finally some conlusions are given.

\section{Whitham hierarchy}

\subsection{Basic definitions and Hamilton-Jacobi problems}

Consider a set of Lax functions $z_\alpha(p,\ti)$ ($\alpha\in
\{\infty,1,...,N\}$) in a complex variable $p$ being formal
Laurent series at $\infty$ and 'finite' punctures $q_i$ on the
Riemann sphere
\begin{align}
\label{lax1} z_\infty &= p + \sum_{l=1}^{\infty}u_{\infty,l}(\ti)p^{-l}\\
\label{lax2} z_i &= u_{i,-1}(\ti)(p-q_i(\ti))^{-1} +
\sum_{l=0}^{\infty}u_{i,l}(\ti)(p-q_i(\ti))^{l}\qquad i\in
{1,...,N},
\end{align}
respectively. In \eqreff{lax1}{lax2} the coefficients $u_{\alpha,l}$
and the poles $q_i$ are smooth dynamical fields depending on the
infinite set of evolution parameters ('times') $\ti=\{t_{\alpha
n}\}$. One of these times is distinguished as a spatial variable:
$t_{\infty 1}=:x$. Times $t_{\alpha n}$ are coupled with the
generating functions $\Omega_{\alpha n}$ defined in
the following way
\begin{align}\label{omega}
\Omega_{\alpha n}(p,\ti):=
 \begin{cases}
\bra{z_\infty^n}_{\bra{\alpha,+}} &\text{ for } \alpha=\infty,\quad n=1,2,3,...\ ,\\
-\bra{z_\alpha^n}_{\bra{\alpha,-}} &\text{ for } \alpha=i,\quad n=1,2,3,...\ ,\\
\ln \bra{p-q_i} &\text{ for } \alpha=i,\quad n=0,
 \end{cases}
\end{align}
where $(\cdot)_{\{\alpha,+\}}$ are the projections onto the
principal parts of Laurent expansions at poles $\infty$~and~$q_i$
such that $(\sum_k a_k p^k)_{(\infty,+)} := \sum_{k\me 0} a_k p^k$
and $(\sum_k a_k(p-q_i)^k)_{(i,-)} := \sum_{k< 0} a_k (p-q_i)^k$,
respectively. Then, the Whitham hierarchy is the set of zero-curvature equations
\begin{equation}
\label{whitham} \pd{\Omega_{\alpha m}}{t_{\beta n}} -
\pd{\Omega_{\beta n}}{t_{\alpha m}} + \pobr{\Omega_{\alpha
m},\Omega_{\beta n}} = 0
\end{equation}
for the following hierarchies of mutually commuting evolution
systems given in the Lax form
\begin{equation}\label{laxh}
\pd{z_\alpha}{t_{\beta m}} = \pobr{\Omega_{\beta m},z_\alpha},
\end{equation}
where the Poisson bracket is the canonical one
\begin{equation}\label{pb1}
\pobr{f,g}:=\pd{f}{p}\pd{g}{x} - \pd{f}{x} \pd{g}{p} .
\end{equation}
Let us use the short notation $t_{\alpha m}\arr t_{\beta n}$ for
the equation, related to fixed times $t_{\alpha m}$ and $t_{\beta
n}$, from the Whitham hierarchy \eqref{whitham}. Notice, that
$\Omega_{\infty 1}=p$ thus $t_{\alpha n}\arr t_{\infty 1}$ is
trivially satisfied and the equations from Lax hierarchies
\eqref{laxh} for $\beta=\infty, m=1$ are the translation symmetries $(z_\alpha)_{t_{\infty 1}}=(z_\alpha)_x$ in the variable $x$ chosen as a spatial one.

The above theory of Whitham hierarchy is well defined due to the
standard approach following from the classical $R$-matrix
formalism \cite{K2,STS,SB}. The Lax hierarchy \eqref{laxh} for
$\alpha=\beta=\infty$ is the well known infinite-field
Benney chain, containing many well
known finite-field reductions being dispersionless limits (see for
example \cite{BS2}) of soliton systems \cite{KO} from the KP
hierarchy.

Examing the bi-Hamiltonian structure of equations \eqref{laxh}
 in the spirit of the classical $R$-matrix theory \cite{SB}, one
observes that hierarchies \eqref{laxh} for different fixed
$\alpha$ and $n\me 1$ form different bi-Hamiltonian hierarchies,
mutually commuting, with respect to the same Poisson tensors.
The symmetries with $n=1$ are the starting symmetries for the
recurrence procedure. In the same way equations for $\alpha=i$ and
$n=0$, constructed for $\Omega=\ln (p-q_i)$, generate other
bi-Hamiltonian hierarchies, which except the first symmetries are
not constructed within the above scheme of the Whitham hierarchy with
generating functions restricted to \eqref{omega}. In this sense the above construction is still incomplete. The appearance of the logarithmic function in
\eqref{omega} could be explained in several ways.
For example in \cite{MMM} all terms \eqref{omega}, including the logarithmic ones, are constructed
by the Lie algebraic splitting of Hamiltonian vector fields.
How to generalize the Whitham hierarchy so that it contains the whole bi-Hamiltonian hierarchies generated by
functions containing logarithmic terms is not clear yet. Restricting
to the particular reductions of the Whitham hierarchy given by
meromorphic Lax functions with only two poles, first $\infty$ and
second 'finite' one, one can manage with this problem proceeding
like in \cite{EY}, where logarithmic terms were represented by a
sum of two infinite series convergent at these poles. For the same
class of Lax functions the more correct approach is given in
\cite{AK}, where the logarithmic terms are constructed using
contour integrals. However, the generalization of such approaches
to the whole Whitham hierarchy makes problems.

\begin{example}\label{ex1}
Consider the part of the Whitham hierarchy involving times:
$t_{\infty 1}=:x$, $t_{\infty 2}=:t$, $t_{i1}=:y_i$ and
$t_{i0}=:s_i$. The above notation will be used in the whole article.
The functions \eqref{omega} related to these times have the form
\begin{equation*}
 \Omega_{\infty 1}=p\qquad \Omega_{\infty 2}=p^2+2w\qquad \Omega_{i1}= - a_i(p-q_i)^{-1}\qquad
 \Omega_{i0}=\ln (p-q_i),
\end{equation*}
where $w:=u_{\infty,1}$, $a_i:=u_{i,-1}$. Then, for $t\arr y_i$
one finds
\begin{equation}\label{eq1}
 (q_i)_t=(2w+q_i^2)_x\qquad (a_i)_t=(2a_i q_i)_x\qquad
 w_{y_i}=(a_i)_x
\end{equation}
the generalized Benney gas system. For $y_i\arr y_j$ and $s_i\arr
y_j$ we have
\begin{equation}\label{eq2}
(q_j)_{y_i}=\bra{\frac{a_i}{q_i-q_j}}_x\qquad
(a_j)_{y_i}=(a_i)_{y_j}=\bra{\frac{a_i a_j}{(q_i-q_j)^2}}_x\qquad
i\neq j
\end{equation}
and
\begin{align}
\label{eq3} &(q_i)_{y_i}=(a_i)_{s_i}\qquad (q_i)_{s_i}=\frac{(a_i)_x}{a_i}\\
\label{eq4} &(q_j)_{s_i}=(q_i)_{s_j}=
\frac{(q_i-q_j)_x}{q_i-q_j}\qquad (a_i)_{s_j}=(q_j)_{y_i}\qquad
i\neq j,
\end{align}
respectively. Eliminating $q_i$ from \eqref{eq3} one obtains the
dispersionless multi-dimensional Toda equation or equivalently the
Boyer-Finley equation, which appears in general relativity theory.
For $s_i\arr t$ one finds
\begin{equation*}
w_{s_i}=(q_i)_x
\end{equation*}
and the first equation from \eqref{eq1}. For $s_i\arr s_j$ we have
the first equation from \eqref{eq4}. The set of equations from
this example has been considered in \cite{MMM}.
\end{example}

Furthermore, there exists a set of pseudo-potential functions
$\s_\alpha (z_\alpha,\ti)$ satisfying a set of the Hamilton-Jacobi
problems
\begin{align}
\label{HJ1} z_\alpha &= z_\alpha \bra{\pr_x \s_\alpha,\ti}\\
\label{HJ2} \pd{\s_\alpha}{t_{\beta n}} &= \Omega_{\beta
n}\bra{\pr_x \s_\alpha,\ti}
\end{align}
and compatibility conditions
\begin{align}\label{cc}
\frac{\pr^2 \s_\gamma}{\pr t_{\alpha m} \pr t_{\beta n}} =
\frac{\pr^2 \s_\gamma}{\pr t_{\beta n} \pr t_{\alpha m}},
\end{align}
since $p=\pr_x\s_\alpha$. The above equations must be correctly
understood, i.e. \eqref{HJ1} are Hamilton-Jacobi equations with
fixed 'energy' levels and therefore $\s_\alpha$ depend on
$z_\alpha$ treated as parameters. Eliminating $\pr_x\s$ from
\eqreff{HJ1}{HJ2} the compatibility conditions \eqref{cc} lead to
the Whitham hierarchy \eqref{whitham} and simultaneously to Lax
hierarchies \eqref{laxh}.

\subsection{Algebraic orbit reductions}

There is important class of finite-field reductions of the Whitham
hierarchy, being the so-called algebraic orbits given by
meromorphic functions $E$ satisfying
\begin{align*}
E= z_\infty^{n_\infty} = z_i^{n_i}\qquad \text{for}\quad
n_\alpha\in \bb{Z}_+ .
\end{align*}
Then, the most general form of $E$ is given by
\begin{equation}\label{mlax}
E = p^{n_\infty} + \sum_{l=0}^{n_{\infty}-2}a_{\infty,l}p^l +
\sum_{i=1}^{N}\sum_{l_i=1}^{n_i}a_{i,l_i}(p-q_i)^{-l_i}
\end{equation}
and then the dynamical fields from functions $z_\alpha$ are given
by polynomials of fields from \eqref{mlax}, i.e. expanding $E$ at $\infty$
$E=z_\infty^{n_\infty}$ and expanding $E$ at $q_i$ $E = z_i^{n_i}$.
The goal of such
reductions of the Whitham hierarchy is the construction of
(1+1)-dimensional integrable finite-field dispersionless systems.
By integrable systems we understand those which have infinite
hierarchy of commuting symmetries. In this case pseudo-potential
functions reduce to one $\s_\alpha(z_\alpha,\ti):= \s(E,\ti)$ and
the set of Lax hierarchies \eqref{laxh} reduces also to one Lax
hierarchy in the form
\begin{equation*}
\pd{E}{t_{\beta m}} = \pobr{\Omega_{\beta m}, E},
\end{equation*}
where the functions $\Omega_{\beta m}$ are generated around poles
of $E$.

A more general theory of meromorphic Lax representations, not only
for the canonical Poisson bracket \eqref{pb1}, including the case
\eqref{mlax} allowing a construction of integrable dispersionless
systems together with multi-Hamiltonian structures is presented in
\cite{SB}. In the same paper, it is also shown that one can construct Lax
hierarchies not only at poles of meromorphic Lax function $E$ but
also at its zeros. Another class of reductions of the Whitham
hierarchy related to algebraic orbits is presented in \cite{GMMA}.

In all examples in this article we will in general present only
the first nontrivial evolution equations from related hierarchies.

\begin{example}\label{ex} The dispersionless AKNS hierarchy.

We will consider the meromorphic Lax function in the form
\begin{equation}\label{dAKNSh}
 E = p + \sum_i a_i (p-q_i)^{-1} = z_\infty = z_i,
\end{equation}
i.e. the case of \eqref{mlax} for $n_\infty=n_i=1$. Then the
system generated by $\Omega_{\infty 2}= p^2 + 2\sum_i a_i$ is
\begin{equation}\label{dAKNS}
\begin{split}
(a_i)_t &= 2(a_i q_i)_x\\
(q_i)_t &= 2\bra{\sum_j a_j + q_i^2}_x.
\end{split}
\end{equation}
This is the dispersionless limit of multi-component AKNS system.
For $\Omega_{k 0}= \ln(p-q_k)$ we have the following equations
\begin{equation*}%\label{dtoda}
\begin{split}
(a_i)_{s_k} &= \bra{\frac{a_i}{q_i-q_k}}_x\\
(q_i)_{s_k} &= \bra{\ln |\, q_i-q_k|}_x\\
(a_k)_{s_k} &= (q_k)_x-\sum_{k\neq k}\bra{\frac{a_k}{q_k-q_k}}_x\\
(q_k)_{s_k} &= \bra{\ln a_k}_x,
\end{split}
\end{equation*}
where $i\neq k$. For $\Omega_{k 1}=-a_k(p-q_k)^{-1}$ one finds
\begin{equation}\label{cos}
\begin{split}
(a_i)_{y_k} &= \bra{\frac{a_i a_k}{(q_i-q_k)^2}}_x\\
(q_i)_{y_k} &= -\bra{\frac{a_k}{q_i-q_k}}_x\\
(a_k)_{y_k} &= (a_k)_x-\sum_{i\neq k}\left(\frac{a_ia_k}{(q_i-q_k)^2}\right)_x\\
(q_k)_{y_k} &= (q_k)_x-\sum_{i\neq
k}\left(\frac{a_i}{q_i-q_k}\right)_x,
\end{split}
\end{equation}
where $i\neq k$. The equations from this example are of course
compatible with the equation from Example \ref{ex1} as in this
case $w=\sum_i a_i$.
\end{example}

\begin{example}
Consider Lax function \eqref{mlax} for $N=1$ with the pole of
first order at $\infty$ and second order pole at $q:=q_1$:
\begin{equation}\label{dbon}
 E = p+u(p-q)^{-1}+ v(p-q)^{-2} = z_\infty = z_1^2.
\end{equation}
Let $y:=y_1$, $s:=s_1$. Then, one finds the following equations
\begin{align}\label{dbon1}
\Omega_{\infty 2} = p^2+2u \qquad \Longrightarrow \qquad
\begin{split}
 u_t &= 2(uq)_x+2v_x\\
 v_t &= 2v_xq+4vq_x\\
 q_t &= 2u_x+2qq_x,
\end{split}
\end{align}
\begin{align}\label{dbon2}
\Omega_{10}= \ln (p-q)\qquad \Longrightarrow \qquad
 \begin{split}
u_s &= q_x\\
v_s &= u_x-\frac{uv_x}{2v}\\
q_s &= \frac{v_x}{2v},
 \end{split}
\end{align}
\begin{align}\label{dbon3}
\Omega_{11} = -\sqrt{v}(p-q)^{-1}\qquad \Longrightarrow \qquad
 \begin{split}
u_y &= (\sqrt{v})_x\\
v_y &= \sqrt{v}\bra{q-\frac{1}{4}\frac{u^2}{v}}_x\\
q_y &= \bra{\frac{u}{2\sqrt{v}}}_x.
 \end{split}
\end{align}
Now $w=u$ from Example \ref{ex1}.
\end{example}

\begin{example}\label{alg21.1}
Let us take now \eqref{mlax} for $N=1$ with $n_{\infty}=2$ and
$n_1=1$, i.e.
\begin{equation*}
E=p^2+u+\frac{a}{p-q} = z_\infty^2 = z_1.
\end{equation*}
In this case $E=\Omega_{\infty 2}-\Omega_{11}$, thus we have that
$E_t=\{\Omega_{\infty2},E\}=\{\Omega_{11},E\}=E_y$ ($y:=y_1$) and
consequently $y$ and $t$ can be identified, i.e. $t \equiv y$. The
system for these times is
\begin{align}\label{exl12}
\Omega_{\infty2}=p^2+u\qquad \Longrightarrow \qquad
 \begin{split}
u_t&= 2a_x\\
a_t&= (2aq)_x\\
q_t&= (q^2+u)_x.
 \end{split}
\end{align}
For the times $\tau:=t_{\infty3}$ and $s:=s_1$ one finds the
following systems
\begin{align}\label{exl13}
\Omega_{\infty 3}= p^3+\frac{3}{2}u p + \frac{3}{2}a\qquad
\Longrightarrow \qquad
 \begin{split}
u_\tau &= \bra{\frac{3}{4}u^2+3aq}_x\\
a_\tau &= \bra{3aq^2+\frac{3}{2}ua}_x\\
q_\tau &= \bra{q^3+\frac{3}{2}uq+\frac{3}{2}a}_x
 \end{split}
\end{align}
and
\begin{align*}%\label{exl11}
\Omega_{10} = \ln (p-q)\qquad \Longrightarrow \qquad
 \begin{split}
u_s &= 2q_x\\
a_s &= (q^2+u)_x\\
q_s &= (\ln a)_x.
 \end{split}
\end{align*}
\end{example}

\subsection{Alternative formulation}\label{sect1}

The choice of the evolution parameter $t_{\infty 1}$ to be
considered as a spatial parameter $x$ is ambiguous. Since
$\Omega_{k 0}=\ln(p-q_k)$ by \eqref{HJ2} the following relation is
valid
\begin{equation}\label{rel}
\pr_x \s_\alpha = \exp \bra{\pr_{s_k} \s_\alpha }+ q_k,
\end{equation}
where $s_k:=t_{k0}$. Now fixing $k$, we can eliminate
$\pr_{s_k}\s_\alpha$ from \eqreff{HJ1}{HJ2} instead of
$\pr_x\s_\alpha$ and consider $s_k$ as a spatial variable. Then,
for $\la_k= \exp(\pr_{s_k}\s_\alpha)$ the Whitham hierarchy
\eqref{whitham} and related Lax hierarchies \eqref{laxh} have the
same form, but the Poisson bracket is given by
\begin{equation}\label{pb2}
\pobr{f,g} = \la_k\bra{\pd{f}{\la_k}\pd{g}{s_k} -
\pd{f}{s_k}\pd{g}{\la_k}} .
\end{equation}
In such a situation the variables $p$ and $\la_k$ should be
considered rather only as auxiliary parameters. On the level of
zero-curvature equations \eqref{whitham} and Lax hierarchies
\eqref{laxh} the above procedure is little bit more complicated.
>From \eqref{rel} it follows that at first, one has to perform
the transformation $\la_k=p-q_k$. Hence, $\pr_p\map \pr_{\la_k}$ and
$\pr_{t_{\alpha n}}\map \pr_{t_{\alpha n}} -(q_k)_{t_{\alpha n}}
\pr_{\la_k}$. Next, all terms in \eqref{whitham} and \eqref{laxh}
containing derivatives with respect to $x$ have to be replaced
using equations $t_{\alpha n}\arr s_k$ for $\Omega_{k 0}=\ln
\la_k$ and \eqref{laxh} for $\beta=k$,~$m=0$. As a result, one
obtains, in a preserved form, the Whitham hierarchy \eqref{whitham}
and related Lax hierarchies \eqref{laxh} for the Poisson bracket
given by \eqref{pb2}. On the level of explicit equations both
formulations give a compatible set of equations. In the case of
explicit equations from Lax hierarchies \eqref{laxh} the above
passage between both formulations relies on the change between
spatial variable $x$ and $s_k$ through use of explicit equations
\eqref{laxh} for $\beta=k, m=0$ or $\beta=\infty, m=1$,
respectively.

Of course in the new auxiliary variable $\la_k$ the Lax functions
\eqreff{lax1}{lax2} take appropriate form of Laurent series. The
"finite" poles $q_i$ are then moved to $q_i-q_k$. Particularly
$z_i$ takes the form of a Laurent series at $0$
\begin{equation*}
 z_i = a_{i,-1} {\la_i}^{-1} +\sum_{l=0}^{\infty}a_{i,l}{\la_i}^l,
\end{equation*}
so the dynamical field $q_i$ in $z_i$ is missed. Then, the
evolution of $q_i$ can be calculated from the representation of
$z_\infty$ as
\begin{equation*}
z_\infty = \la_k + q_k +
\sum_{l=1}^{\infty}a_{\infty,l}\la_k^{-l}.
\end{equation*}

In fact, we can use all auxiliary variables $p$ and $\la_i$ (for
all $i$) simultaneously. This formulation of the Whitham hierarchy
is correct if all equations (relations) are understood on the
level of the Hamiltonian-Jacobi problems, i.e. $p\equiv
\pr_x\s_\alpha$, $\la_i\equiv \exp(\pr_{s_i}\s_\alpha)$ are not
considered as independent variables but as shortened notation.
Then, the Poisson bracket is given in the following general form
\begin{equation*}
\pobr{f,g}=  \pd{f}{p}\pd{g}{x} - \pd{f}{x} \pd{g}{p}  + \sum_i
\la_i\bra{\pd{f}{\la_i}\pd{g}{s_i} - \pd{f}{s_i}\pd{g}{\la_i}}.
\end{equation*}
In this case one has to first evaluate the Poisson bracket and
then use the relation $p\equiv \la_i+q_i\equiv \la_j+q_j$ to draw
consistent equations from \eqref{whitham} or \eqref{laxh}.
However, one has to proceed carefully in order to not lose
the evolution of the fields $q_i$.

\begin{example}
With the use of $\la_i$, the functions \eqref{omega}
related to the set of evolution parameters $x$, $t$, $s_i$ and
$y_i$ are given in the form
\begin{equation*}
 \Omega_{\infty 1}= \la_k + q_k\qquad \Omega_{\infty 2}=\la_k^2+2q_k\la_k+q_k^2+2w\qquad
 \Omega_{i1}= - a_i(\la_i)^{-1} \qquad \Omega_{i0}=\ln \la_i,
\end{equation*}
where $k$ is arbitrary from allowed values. Then, one obtains a
set of equations being consistent with the set of equations from
Example \ref{ex1}.
\end{example}

\subsection{Finite-field reductions}

The meromorphic reductions \eqref{mlax} can be represented using
variables $\la_i = p-q_i$ simultaneously in the form
\begin{equation}\label{ml}
E = \la_k^{n_\infty} + u\la_k^{n_\infty-1} +
\sum_{l=0}^{n_{\infty}-2}u_{\infty,l}\la_k +
\sum_{i=1}^{N}\sum_{l_i=1}^{n_i}a_{i,l_i} \la_i^{-l_i},
\end{equation}
where $k$ is arbitrary and $u=n_\infty q_k$. However, as we are
interested in the construction of (1+1)-dimensional equations we
should use the form of $E$ only with one auxiliary parameter
$\la_k$ for a fixed $k$. In this case, $\Omega_{k 0}= \ln \la_k$
leads to the trivially satisfied equations related to the
translational symmetry, this time, in the variable $s_k$.
Therefore, the equation from the hierarchy \eqref{laxh} for $\alpha=k$
which was earlier calculated for $\Omega_{k 0} =\ln (p-q_k)$ can
now be calculated for $\Omega_{\infty 1} = \la_k+q_k$ but in the
reversed form, i.e. with $x$ as an evolution parameter.

\begin{example}Multi-component dispersionless Toda hierarchy.

In the auxiliary parameter $\la_k = p-q_k$ the Lax function
\eqref{dAKNSh} takes the form
\begin{equation*}
 E = \la_k + q_k + a_k\la_k + \sum_{i\neq k} a_i (\la_k + q_k-q_i)^{-1}.
\end{equation*}
The first nontrivial system is a counterpart of translational symmetry
in $x$ for \eqref{dAKNSh}
\begin{align}\label{dtoda2}
\Omega_{\infty 1} = \la_k + q_k \qquad \Longrightarrow \qquad
\begin{split}
(a_i)_x &= (q_i-q_k) (a_i)_{s_k} + a_i(q_k)_{s_k}\\
(q_i)_x &= (q_i-q_k) (q_i)_{s_k} + \sum_j (a_j)_{s_k}.
\end{split}
\end{align}
This is the multi-component dispersionless Toda equation. The
counterpart of dispersionless AKNS system \eqref{dAKNS} is only
the next system from the hierarchy calculated at $\infty$. For
$\Omega_{k1} = -a_k \la_k^{-1}$ one obtains a counterpart of the
system \eqref{cos}, which can be also calculated directly from
\eqref{cos} with the use of \eqref{dtoda2},
\begin{equation*}
\begin{split}
(a_i)_{y_k} &= \frac{(a_i)_{s_k} a_k - a_i (a_k)_{s_i}
}{q_i-q_k} \qquad i\neq k\\
(a_k)_{y_k} &= a_k (q_k)_{s_k} - \sum_{j\neq k}(a_j)_{y_k}\\
(q_i)_{y_k} &= (a_k)_{s_i}.
\end{split}
\end{equation*}
As a result, the dispersionless AKNS hierarchy from Example
\ref{ex}, with $x$ as a spatial variable, becomes a dispersionless
multi-component Toda hierarchy with the spatial variable $s_k$.
\end{example}

\begin{example}
In $\la := \la_1 = p - q$ the Lax operator \eqref{dbon} has the
form
\begin{equation}\label{dbon0}
 E = \la + q + u\la^{-1} +v\la^{-2}
\end{equation}
and the first nontrivial equation becomes
\begin{align}\label{aaa}
\Omega_{\infty 1} = \la + q \qquad \Longrightarrow \qquad
 \begin{split}
u_x &= uq_s+v_s\\
v_x &= 2vq_s\\
q_x &= u_s .
 \end{split}
\end{align}
The counterpart of \eqref{dbon2} is only the next system from
the hierarchy. Then,
\begin{align*}
\Omega_{11}=-\sqrt{v}\la^{-1}\qquad \Longrightarrow \qquad
 \begin{split}
u_y &= \sqrt{v}q_s\\
v_y &= v\bra{\frac{u}{\sqrt{v}}}_s\\
q_y &= (\sqrt{v})_s
 \end{split}
\end{align*}
is the counterpart of \eqref{dbon3}.
\end{example}

\begin{example}\label{alg21.2}
In terms of $\lambda:=p-q$ we can also rewrite the algebraic orbit of
Example \ref{alg21.1} as:
\begin{equation*}
 E=\la^2+2q\la+q^2+u+a\la^{-1}.
\end{equation*}
So, we have
\begin{align}\label{exq1}
\Omega_{\infty 1} = \la + q\qquad \Longrightarrow \qquad
 \begin{split}
u_x &= a_s-qu_s\\
a_x &= aq_s\\
q_x &= \frac{u_s}{2}.
 \end{split}
\end{align}
As before $E=\Omega_{\infty 2} - \Omega_{11}$. The counterpart of
\eqref{exl12} is
\begin{align}\label{exq2}
\Omega_{\infty 2} = \la^2+2q\la+q^2+u\qquad \Longrightarrow \qquad
 \begin{split}
u_t &= 2aq_s\\
a_t &= a(q^2+u)_s\\
q_t &= a_s.
 \end{split}
\end{align}
%and for $\Omega_{\infty 3} =
%\la^3+3q\la^2+\bra{3q^2+\frac{3}{2}u}\la+q^3+\frac{3}{2}qu+\frac{3}{2}a$
%\begin{align*}
% \begin{split}
%u_{\tau} &= \frac{3}{2}(ua)_s+3aqq_s-\frac{3}{2}quu_s\\
%a_{\tau} &= a\bra{\frac{3}{2}a+\frac{3}{2}uq+q^3}_s\\
%q_{\tau} &= \frac{3}{8}\bra{4qa+u^2}_s.
% \end{split}
%\end{align*}
\end{example}

\section{Dispersionful Whitham hierarchy}

\subsection{Pseudo-differential operators and auxiliary linear problems}

In this section we will use the same set of evolution parameters
$t_{\alpha n}$ and the same notation $x:=t_{\infty 1}$, $s_i:=t_{i
0}$ and $y_i:=t_{i 1}$ as in the previous section.

It is well known that the Benney momentum chain can be obtained
from the infinite-field KP hierarchy in the quasi-classical limit,
which suggests that the dispersive counterpart of the Whitham
hierarchy can be constructed by means of pseudo-differentials
operators $\pr:=\pr_x$ satisfying the generalized Leibniz rule
\begin{equation}\label{lebniz}
\pr^m u(x) = \sum_{n\me 0} \binom{m}{n} u(x)_{nx} \pr^{m-n},
\end{equation}
where $\binom{m}{n}=(-1)^n\binom{-m+n-1}{n}$ for $m<0$. From
\eqref{lebniz} it follows that
\begin{align*}
(\pr-v) u &= u(\pr-v) +u_x\\
(\pr-v)^{-1}u &= u(\pr-v)^{-1}-(\pr-v)^{-1}u_x(\pr-v)^{-1} \\
&= u(\pr-v)^{-1} - u_x(\pr-v)^{-2}+u_{2x}(\pr-v)^{-3} - ...\ ,
\end{align*}
where $v$ can be equal zero. Some other useful formulae, needed in
further calculations, are
\begin{align*}
&\pr(\pr-v)^{-1} = 1+v(\pr-v)^{-1}\\
&(\pr-v)^{-1}\pr = 1+(\pr-v)^{-1}v\\
&(\pr-v)^{-1}u=u\bra{\pr-v+\tfrac{u_x}{u}}^{-1}\\
&\brac{(\pr-v)^{-1}}_t =(\pr-v)^{-1}v_t(\pr-v)^{-1}.
\end{align*}

Let us consider the following algebras, generated respectively by
elements $\pr,\pr^{-1}$ and $(\pr-q_i),(\pr-q_i)^{-1}$,
\begin{align*}
\alg_\infty &= \alg_{(\infty,+)}\oplus \alg_{(\infty,-)}=
\pobr{\sum_{l\me 0} a_l\pr^l}\oplus
\pobr{\sum_{l< 0} a_l\pr^l}\\
\alg_i &= \alg_{(i,+)}\oplus \alg_{(i,-)}= \pobr{\sum_{l\me 0}
a_l(\pr-q_i)^l}\oplus \pobr{\sum_{l< 0} a_l(\pr-q_i)^l},
\end{align*}
in which the operation, with respect to the above rules, is
associative. The Lie algebra structure is defined through
the commutator $[A,B]= AB-BA$, where $A,B\in \alg_\alpha$, and these
algebras are decomposed into Lie subalgebras. We would like to
have on the 'quantum' level of pseudo-differential operators a
theory parallel to the complex calculus of meromorphic functions,
but after quantization we do not have to our disposal such a
crucial tool as expansion into Taylor series and the clear-cut
notion of poles for pseudo-differential operators. However, it is
possible to construct formal rules of expansions which are similar
to the expansions for meromorphic functions, i.e. $\alg_\infty,
\alg_j \subset \alg_{(i,+)}$ for $j\neq i$ and $\alg_{(i,\pm)}
\subset \alg_{(\infty,\pm)}$, where $\mathfrak{a}\subset
\mathfrak{b}$ means that all elements from $\mathfrak{a}$ can be
expanded as a series from $\mathfrak{b}$. Hence, the reasoning is
analogous to the one on the Whitham hierarchy level. This is a
crucial fact for the consistency of the equations \eqref{lh} and
\eqref{zerc}. The above scheme will be considered in detail in the
forthcoming article \cite{S}.

We define the infinite-field Lax operators in the form
\begin{align}
\label{Lax1} L_\infty &= \pr + \sum_{l=1}^{\infty}u_{\infty,l}(\ti)\pr^{-l}\\
\label{Lax2} L_i &= u_{i,-1}(\ti) (\pr-q_i(\ti))^{-1} +
\sum_{l=0}^{\infty}u_{i,l}(\ti) (\pr-q_i(\ti))^{l}\qquad i\in
{1,...,N},
\end{align}
where all coefficients depend on the same set of evolution
parameters as before. The first one $L_\infty$ defines the
well-known KP hierarchy. Then
\begin{align}\label{om}
\Omega_{\alpha n}(\ti):=
\begin{cases}
\bra{L_\infty^n}_{\bra{\infty,+}} \qquad\text{ for } \alpha=\infty\\
-\bra{L_i^n}_{\bra{i,-}} \qquad\text{ for } \alpha=i,
\end{cases}
\end{align}
where $n=1,2,3,...$. Due to the lack of a good counterpart of
logarithmic function, formulated by means of pseudo-differential
operators, we omitted the case related to $\Omega_{i0}$ in
\eqref{omega} with the natural logarithm. It will be considered
separately.

Now we can define a set of auxiliary linear equations
\begin{align}
\label{lin1} L_\alpha \Psi_\alpha &= z_\alpha \Psi_\alpha\\
\label{lin2} \pd{\Psi_\alpha}{t_{\beta n}} &= \Omega_{\beta n}
\Psi_\alpha.
\end{align}
for some eigenfunctions $\Psi_\alpha(\ti)$ and eigenvalues
$z_\alpha$. Then, the compatibility conditions between
\eqref{lin1} and \eqref{lin2} lead to the Lax hierarchies
\begin{equation}\label{lh}
\brac{\pr_{t_{\beta n}}-\Omega_{\beta n},L_\alpha}\Psi_\alpha=0.
\end{equation}
The compatibility conditions among equations \eqref{lin2} lead to
the zero-curvature equations
\begin{equation}
\label{zerc} \brac{\pr_{t_{\alpha m}}-\Omega_{\alpha
m},\pr_{t_{\beta n}}-\Omega_{\beta n}}\Psi_\gamma = 0
\end{equation}
of Lax hierarchies \eqref{lh}. They are written as operation on
eigenfunctions as this form guarantees that they are invariant
with respect to multi-component Toda-like formulation presented in
Section~\ref{sect}. Still we have to argue that they are well
defined. One can show that equations \eqref{lh} are
self-consistent, i.e. Lax operators \eqreff{Lax1}{Lax2} are well
normalized and the right- and left-hand sides of Lax hierarchies
can be written in the same form. The related $R$-matrices are
defined through decomposition of Lie algebras $\alg_\alpha$ and
importantly they are mutually consistent due to the above rules of
expansion. The mutual commutativity between symmetries from fixed
Lax hierarchy \eqref{lh} follows now from the scheme of $R$-matrix
theory \cite{STS,BS}. We will use the notation $t_{\alpha m}\arr
t_{\beta n}$ for the related equation from the zero-curvature
hierarchy \eqref{zerc}. Notice, that $\Omega_{\infty,1}=\pr$.
Hence, $t_{\alpha n}\arr t_{\infty 1}$ is trivially satisfied and
in the representation of pseudo-differential operators
$(L_\alpha)_{t_{\infty 1}}=(L_\alpha)_x$.

The so-called quasi-classical limit of linear equations
\eqreff{lin1}{lin2} leads to the Hamiltonian-Jacobi equations
\eqreff{HJ1}{HJ2} for $n\neq 0$. To see that, at first one has to
perform the transformation
\begin{align*}
t_{\alpha n} \map \frac{1}{\hk} t_{\alpha n} \Longrightarrow
\pr_{t_{\alpha n}}\map \hk \pr_{t_{\alpha n}},
\end{align*}
where $\hk$ is treated as a deformation parameter. Next, assuming
the WKB form of the eigenfunctions and that dynamical fields have
quasi-classical counterparts, i.e.
\begin{align*}
&\Psi_\alpha = \exp \bra{\frac{1}{\hk}\s_\alpha + O(\hk)}\\
&u_{\alpha,l}\bra{\tfrac{1}{\hk}\ti} =
\widetilde{u}_{\alpha,l}(\ti) + O(\hk),
\end{align*}
one takes the limit $\hk \arrow 0$. Then, the fields
$\widetilde{u}_{\alpha,l}(\ti)$ can be identified with those from
Lax functions \eqreff{lax1}{lax2}. Hence, the zero-curvature
equations \eqref{zerc} can be considered as dispersive analogues
of the Whitham equations \eqref{whitham} without the case related
to $n=0$. However, such an approach with the use of the formalism
of pseudo-differential operators has fundamental disadvantages.
The first one is that the coefficients of $\Omega_{i,n}$ for $n\me
2$ do not have 'limited' or 'compact' form, which means that they
are differential polynomials in all, infinitely many, coefficients
from Lax operators $L_i$. It makes particular problems when
one tries to define counterparts of algebraic orbits for the
original Whitham hierarchy.

\begin{example}\label{exa}
Consider equations from the dispersive Whitham hierarchy related
to the times: $t_{\infty1}=:x$, $t_{\infty2}=:t$, $t_{i1}=:y_i$ for
which the related operators \eqref{om} have the form
\begin{equation*}
\Omega_{\infty 1}\Psi_\alpha = \pr\Psi_\alpha\qquad \Omega_{\infty
2}\Psi_\alpha=\bra{\pr^2+2w}\Psi_\alpha\qquad
\Omega_{i1}\Psi_\alpha= - a_i(\pr-q_i)^{-1}\Psi_\alpha,
\end{equation*}
where $w:=u_{\infty,1}$, $a_i:=u_{i,-1}$. Equations for
$\Omega_{\infty 1}$ are satisfied trivially. Then, for $t\arr y_i$
one finds
\begin{equation}\label{e1}
(a_i)_t = (a_i)_{2 x} + 2(a_i q_i)_x\qquad (q_i)_t = -(q_i)_{2x}
+ 2q_i(q_i)_x+ 2w_x\qquad w_{y_i}=(a_i)_x,
\end{equation}
i.e. the dispersive counterpart of the generalized Benney gas
system \eqref{eq1}. To calculate the equation related to $y_i\arr
y_j$ it is better to rewrite $\Omega_{i1}$ in the form
$\Omega_{i1}\Psi_\alpha=-\psi_i\pr^{-1}\phi_i\Psi_\alpha$, where
$a_i = \psi_i \phi_i$ and $q_i = -(\ln \phi_i)_x$. Then, the
dispersive versions of \eqref{eq2} are
\begin{equation*}
(\psi_j)_{y_i}=-\psi_i \pr_x^{-1}(\psi_j \phi_i)\qquad
(\phi_j)_{y_i}=-\phi_i \pr_x^{-1}(\psi_i \phi_j),
\end{equation*}
where $\pr_x^{-1}$ means the formal integration operation which
cannot be confused with $\pr^{-1}$. With the use of $\psi_i$ and
$\phi_i$ the system \eqref{e1} becomes
\begin{equation*}
(\psi_i)_t=(\psi_i)_{2x}+2w\psi_i\qquad
(\phi_i)_t=-(\phi_i)_{2x}-2w\phi_i\qquad w_{y_i}=(\psi_i \phi_i)_x
.
\end{equation*}
\end{example}

\subsection{Finite-field reductions}

After quantization we would like to construct reductions of the
above dispersive Whitham hierarchy being counterparts to algebraic
orbits for the Whitham hierarchy, i.e. we are looking for a Lax
operator $\lo$ satisfying the following linear equation
\begin{equation}\label{algo}
\lo \Psi = L_\infty^{n_\infty} \Psi = L_i^{n_i}\Psi = E \Psi\qquad
n_\infty, n_i \me 1,
\end{equation}
where now the eigenfunctions $\Psi_\alpha$ reduce to one $\Psi$
and we have only one eigenvalue $E$. The related Lax hierarchy
takes the form
\begin{equation}\label{lhful}
\brac{\pr_{t_{\beta n}}-\Omega_{\beta n}, \lo}\Psi=0.
\end{equation}
The disadvantage mentioned above practically makes it impossible
to construct $\lo$ related to $n_i\me 2$ with the use of
pseudo-differential operators. For $n_i=1$ this operator can be
defined as
\begin{equation}\label{orlov}
\lo \Psi= \bra{\pr^{n_\infty} +
\sum_{l=0}^{n_{\infty}-2}u_{l}\pr^l +
\sum_{i=1}^{N}a_{i}(\pr-q_i)^{-1}}\Psi .
\end{equation}
It is useful to replace $a_{i}(\pr-q_i)^{-1}$ by the operator
$\psi_i\pr^{-1}\phi_i$, where $a_i = \psi_i \phi_i$ and $q_i =
-(\ln \phi_i)_x$. Then, \eqref{orlov} takes the form
\begin{equation}\label{orlov2}
\lo \Psi= \bra{\pr^{n_\infty} +
\sum_{l=0}^{n_{\infty}-2}u_{l}\pr^l +
\sum_{i=1}^{N}\psi_i\pr^{-1}\phi_i}\Psi .
\end{equation}
The Lax operators in the form \eqref{orlov2} have been considered
in \cite{OS} as reductions of the KP hierarchy, i.e. only for
$\alpha=\infty$. In a more general context, i.e. for all $\alpha$,
the theory of such operators together with Hamiltonian structures
is presented in \cite{EOR}. The fields $\psi_i$, $\phi_i$ are the
so-called source terms as $\psi_i$ and $\phi_i$ are eigenfunctions
and adjoint-eigenfunctions of the Lax hierarchy \eqref{lhful},
respectively, i.e. $\psi_i$ satisfy linear equations \eqref{lin2}
on eigenfunction $\Psi$.

\begin{example}The multi-component AKNS hierarchy.

Consider the case $n_\infty=1$ of \eqref{orlov} or equivalently
\eqref{orlov2}, that is
\begin{equation}\label{at}
\lo \Psi = \bra{\pr + \sum_i a_i(\pr-q_i)^{-1}} \Psi =
 \bra{\pr + \sum_i \psi_i\pr^{-1}\phi_i} \Psi.
\end{equation}
Then for $\Omega_{\infty 2} \Psi = (\pr^2 + 2\sum_j a_j)\Psi =
(\pr^2 + 2\sum_j \psi_j \phi_j)\Psi$ one calculates the
multi-component AKNS system in a form similar to its
dispersionless limit \eqref{dAKNS}
\begin{equation*}
\begin{split}
(a_i)_t &= (a_i)_{2 x} + 2(a_i q_i)_x\\
(q_i)_t &= -(q_i)_{2x} + 2\sum_j(a_j)_x + 2q_i(q_i)_x
\end{split}
\end{equation*}
or in the representation with source fields
\begin{equation*}
\begin{split}
(\psi_i)_t &= (\psi_i)_{2 x} + 2\psi_i\sum_j \psi_j\phi_j\\
(\phi_i)_t &= -(\phi_i)_{2x} - 2\phi_i\sum_j \psi_j\phi_j.
\end{split}
\end{equation*}
For $\Omega_{k1}\Psi=-\psi_k\pr^{-1}\phi_k\Psi$ one finds
a dispersive counterpart of \eqref{cos}
\begin{equation}\label{AKNS2}
\begin{split}
(\psi_i)_{y_k} &= -\psi_k\pr_x^{-1}(\psi_i\phi_k)\\
(\phi_i)_{y_k} &= -\phi_k\pr_x^{-1}(\psi_k\phi_i)\\
(\psi_k)_{y_k} &= (\psi_k)_x+\sum_{j\neq k}\psi_j
\pr_x^{-1}(\psi_k\phi_j)\\
(\phi_k)_{y_k} &= (\phi_k)_x+\sum_{j\neq k}\phi_j
\pr_x^{-1}(\psi_j\phi_k),
\end{split}
\end{equation}
where $i\neq k$. Obviously the above systems are compatible with
the equations from Example \ref{exa} as $w=2\sum_i a_i$.
\end{example}

\begin{example}\label{alg21.3}
Let us take dispersionful analogue of the Lax function from
Example \ref{alg21.1}, then we have
\begin{equation*}
  \lo \Psi = \bra{\partial^2+u+a(\partial-q)^{-1}}\Psi.
\end{equation*}
We have that $\lo\Psi= \bra{\Omega_{\infty 2}-\Omega_{11}}\Psi$.
So, as before $\lo_t=\lo_y$, i.e. $t\equiv y$. Thus,
\begin{align*}
\Omega_{\infty 2}\Psi = \bra{\pr^2+u}\Psi \qquad \Longrightarrow
\qquad
\begin{split}
u_t &= 2a_x\\
a_t &= (a_x+2aq)_x\\
q_t &= (-q_x+q^2+u)_x,
\end{split}
\end{align*}
which is the dispersionful version of \eqref{exl12}. For
$\Omega_{\infty3}\Psi=\bra{\pr^3+\frac{3}{2}u\pr+\frac{3}{4}(u_x+2a)}\Psi$
one obtains the dispersionful version of \eqref{exl13}
\begin{align*}%\label{exf13}
u_{\tau} &= \bra{\frac{1}{4}u_{2x}+\frac{3}{2}a_x+\frac{3}{4}u^2+3aq}_x\\
a_{\tau} &= \bra{a_{2x}+3a_xq+3aq^2+\frac{3}{2}ua}_x\\
q_{\tau} &=
\bra{q_{2x}-3qq_x-\frac{3}{4}u_x+q^3+\frac{3}{2}uq+\frac{3}{2}a}_x.
\end{align*}
\end{example}

\subsection{Multi-component Toda-like formulation}\label{sect}

In \cite{BS} a unified approach to systematic construction of
field and lattice soliton systems was presented. It was shown
that Lie algebras of pseudo-differential operators can be
considered as Weyl-Moyal-like deformations of algebras with
Poisson structure given by canonical Poisson bracket \eqref{pb1}.
The Weyl-Moyal-like deformation is a special case of the
deformation quantization. On the other hand, Lie algebras of shift
operators in the same procedure can be considered as deformations
of algebras with Poisson bracket in the form \eqref{pb2}, which
suggests that shift operators and similar relation to \eqref{t2}
can be helpful in solving problems appearing in the
pseudo-differential representation of dispersionful analogue of
the Whitham hierarchy. Hence, instead of constructing the operator
counterpart of the logarithm functions $\Omega_{i0}$ we will use,
following \cite{T}, the 'quantum' analogue of the relation
\eqref{rel}, given at this moment {\it ad-hoc} as
\begin{equation}\label{qrel}
\pd{\Psi_\alpha}{x} = \bra{e^{\pr_{s_k}}+ q_k(s_k+1)} \Psi_\alpha,
\end{equation}
where $e^{\pr_s}\Psi = \sum_{l=0}^\infty \frac{1}{n!}\pr_s^n \Psi$
is the contracted formula for Taylor expansion. Then, the
quasi-classical limit of \eqref{qrel} is exactly the relation
\eqref{rel}. Let us denote by $\e_k:=e^{\pr_{s_k}}$ the shift
operators that when acting on dynamical fields satisfy
\begin{equation*}
 \e_k^m u(s_k) = u(s_k+m) \e_k^m\qquad m\in \Z.
\end{equation*}
The field $q_k$ has shifted the argument in \eqref{qrel}, so then it follows that
\begin{equation}\label{qrel2}
\e_k \Psi_\alpha = (\pr - q_k(s_k+1)) \Psi_\alpha \Longrightarrow
\e_k^{-1} \Psi_\alpha = (\pr - q_k)^{-1} \Psi_\alpha.
\end{equation}
>From \eqref{qrel2} we have also the relation between different
shift operators
\begin{equation}\label{qrel3}
\e_i \Psi_\alpha = \bra{\e_k + q_k(s_k+1) - q_i(s_i+1)}
\Psi_\alpha.
\end{equation}
It is important to remember that the above rules are valid only when
acting on eigenfunctions $\Psi_\alpha$. Other relations which may
be useful are
\begin{align*}
&\e_k (\e_k-v)^{-1} = v(\e_k-v)^{-1}+1\\
&(\e_k-v)^{-1}\e_k =(\e_k-v)^{-1}v+1\\
&(\e_k-v)^{-1}u =u(s_k-1)(\e_k-v)^{-1}-(\e_k-v)^{-1}v\brac{u-u(s_k-1)}(\e_k-v)^{-1}\\
&(\e_k-v)^{-1}u = u(s_k-1)\bra{\e_k-v\tfrac{u(s_k-1)}{u}}^{-1}\\
&\brac{(\e_k-v)^{-1}}_t=(\e_k-v)^{-1}v_t(\e_k-v)^{-1}.
\end{align*}

We can add the linear relations \eqref{qrel} to the auxiliary
linear equations \eqref{lin2} and complete the set of
zero-curvature equations \eqref{zerc} by
\begin{align}
\label{r1} &s_i\arr t_{\alpha m}:\qquad \brac{\pr - \e_i -
q_i(s_i+1),
\pr_{t_{\alpha m}}-\Omega_{\alpha m}}\Psi_\gamma = 0\\
\label{r2} &s_i\arr s_j:\qquad \brac{\pr - \e_i - q_i(s_i+1), \pr
- \e_j - q_j(s_j+1)}\Psi_\gamma = 0.
\end{align}
Adequately the Lax hierarchies \eqref{lh} are completed by
\begin{equation*}
\brac{\pr - \e_i - q_i{(s_i+1)}, L_\alpha}\Psi_\alpha=0.
\end{equation*}
They differ from the rest of equations belonging to the
dispersionful Whitham hierarchy as they are partially lattice
systems in variables $s_i$. With supplementary equations \eqreff{r1}{r2} the
dispersionful Whitham hierarchy is in one to one corespondence to the
original one. These equations are well defined as using the
relation \eqref{qrel} we can pass from pseudo-differential
operators to shift ones and formulate the dispersive analogue of
the Whitham hierarchy entirely by means of shift operators. This is a
counterpart of the alternative formulation performed in the
Section~\ref{sect1}.

We define the following Lie algebras of shift operators
\begin{equation*}
\mg_i = \mg_{(i,+)}\oplus \mg_{(i,-)}= \pobr{\sum_{l\me 0} a_l
\e_i^l}\oplus \pobr{\sum_{l< 0} a_l \e_i^l}\qquad i=1,2,...,N,
\end{equation*}
that are decomposed into Lie subalgebras with respect to
the commutator. The different Lie algebras $\mg_i$ are mutually connected through the relations \eqref{qrel2} and \eqref{qrel3} due to which also
the rules of expansion can be formulated. Thus, the algebras of pseudo-differential and shift operators act compatibly on the eigenfunctions $\Psi_\alpha$, i.e.
$\alg_{(i,\pm)} \Psi_\alpha = \mg_{(i,\pm)} \Psi_\alpha$ and
$\alg_{(\infty,\pm)} \Psi_\alpha = \mg_{(i,\pm)} \Psi_\alpha$, for
arbitrary $i$. Hence, the rules of expansion and $R$-matrices
related to decomposition of Lie algebras are invariant under the passage
between both alternative formulations. Moreover, the form of Lax hierarchies \eqref{lh} and zero-curvature equations \eqref{zerc} is preserved. Then, the set of linear equations \eqref{lin1} following from \eqreff{Lax1}{Lax2} is given by
\begin{equation}\label{sato}
\begin{split}
L_\infty \Psi_\alpha &= \bra{\e_k + q_k(s_k+1) + \sum_{l=1}^\infty
a_{\infty,l}\e_k^{-l}}\Psi_\alpha = z_\infty \Psi_\alpha\\
L_i \Psi_\alpha &= \bra{a_{i,-1}\e_i^{-1} +
\sum_{l=0}^{\infty}a_{i,l}\e_i^l} \Psi_\alpha = z_i \Psi_\alpha
\end{split}
\end{equation}
and the definition of generating operators \eqref{om} remain unchanged.
Thus, equation \eqref{qrel} follows from \eqref{lin2} for $\beta=\infty,n=1$. The zero-curvature equations \eqref{zerc} together with Lax hierarchies \eqref{lh} defined for Lax operators \eqref{sato} can be extracted from the 'charged' multi-component KP hierarchy \cite{T}. As we are using simultanously
shift operators $\e_i$ with respect to different $s_i$ we call such a formulation of the dispersionful Whitham hierarchy as the multi-component Toda-like.

In the above formulation there is still missing the 'quantum' counterpart of $\Omega_{k0}=\ln~\la_k$ from Section~\ref{sect1}. The natural logarithm of the shift operator can be defined in the way $\ln~\e_k:=\Diff{\e_k^\alpha}{\alpha}{\alpha=0}$,
where $\alpha$ is arbitrary. Over the logarithm $\ln \e_k$, for fixed $k$, the algebra $\mg_k$ can be extended in such a way that the Lie algebra splitting is be preserved. Thus, we have
\begin{equation*}
 \pd{\Psi_\alpha}{t_{k0}} = \Omega_{k0}\Psi_\alpha = \ln \e_k \Psi_\alpha
= \Diff{e^{\alpha\pr_{s_k}}} {\alpha}{\alpha=0}\Psi_\alpha
= \left.\pr_{s_k}e^{\alpha\pr_{s_k}}\right|_{\alpha=0}\Psi_\alpha
= \pr_{s_k}\Psi_\alpha.
\end{equation*}
Hence $t_{k0}$ can be identified with $s_k$ and the related equations from the dispersionful Whitham hierarchy, in the multi-component Toda-like formulation, are trivially satisfied as it should be. In the case of Lax hierarchies one obtains the translational symmetries $(L_\alpha)_{t_{k0}} =
(L_\alpha)_{s_k}$. As before, in the case of the original Whitham hierarchy, considering the bi-Hamiltonian formulation one would see that these translational symmetries generates higher order symmetries that are not included in our construction. The way of construction of the whole hierarchy generated by logarithmic operators, defined by means of dressing operators, for the two-field lattice Toda system is presented in \cite{CDZ}.

Notice that $\ln \e_k \Psi_\alpha \neq \ln \bra{\pr-q_k(s_k+1)}\Psi_\alpha$
as $\e_k^\alpha\Psi_\alpha \neq \bra{\pr-q_k(s_k+1)}^\alpha\Psi_\alpha$.
Besides, the action of $\ln \bra{\pr-q(s_k+1)}$ on eigenfunctions $\Psi_\alpha$ would not be well defined. Nevertheless, one would like to have the formulation of the above logarithms by means of pseudo-differential operators. To calculate it one should use a formula similar to \eqref{form}, but the calculations are not straightforward anymore.

\begin{example} To have a full dispersive counterpart of
Example \eqref{ex1} we have to complete the set of equations from
Example \eqref{exa} and calculate systems with the evolution
parameters $s_k$. We will use the related operators \eqref{om}
from auxiliary linear equations \eqref{lin2} in the form
\begin{align*}
&\bra{\Psi_\alpha}_x= \bra{\e_k+q_k(s_k+1)}\Psi_\alpha\qquad
\bra{\Psi_\alpha}_{y_i}= \bra{-a_i\e_i^{-1}}\Psi_\alpha\\
&\bra{\Psi_\alpha}_t =  \bra{\e_k^2+\bra{q_k(s_k+2)+q_k(s_k+1)}\e_k+(q_k(s_k+1))^2+a(s_k+1)+a}
\Psi_\alpha.
\end{align*}
Then, for $t\arr y_i$ and $t\arr s_i$ one finds equations
\eqref{e1} and
\begin{equation*}
 (q_i)_x = w-w(s_i-1).
\end{equation*}
For $s_i\arr s_j$ and $y_i\arr y_j$ one finds systems
\begin{align}
\label{q1}
&q_j-q_j(s_i-1)=q_i-q_i(s_j-1)= \bra{\ln |q_i-q_j|}_x\\
\notag %\label{q2}
&(a_j)_{y_i}=(a_i)_{y_j}=\frac{a_ia_j(s_i-1)-a_i(s_j-1)a_j}{q_i(s_j-1)-q_j(s_i-1)},
\end{align}
respectively. Finally, for $s_i\arr y_j$ we have equations
\begin{equation*}
(q_i)_{y_j}=a_j-a_j(s_i-1)\qquad  (a_i)_x=a_i(q_i(s_i+1)-q_i)
\end{equation*}
from which eliminating $q_i$ one finds the multi-dimensional Toda
equation.
\end{example}

\subsection{Finite-field reductions}

Now, in the representation of shift operators we are able
to construct finite field reductions \eqref{algo} for arbitrary
$n_\alpha>0$ given by a Lax operator $\lo$ in the form
\begin{equation}\label{rat}
\lo \Psi = \bra{\e_k^{n_\infty} + u\e_k^{n_\infty-1} +
\sum_{l=0}^{n_{\infty}-2}u_l \e_k^l +
\sum_{i=1}^{N}\sum_{l_i=1}^{n_i}a_{i,l_i}\e_i^{-l_i}} \Psi = E
\Psi
\end{equation}
being a dispersive counterpart of \eqref{ml}, where now
$u=\sum_{j=1}^{n_\infty}q_k(s_k+j)$. Notice that some terms
of \eqref{om} can have coefficients that are nonlocal on lattice, see
Example \ref{exa2}. However, as we are interested in the
construction of (1+1)-dimensional equations we have to show how to
represent \eqref{rat} only with $\e_k$ for fixed $k$. The equation
\eqref{q1} belongs to the zero-cutvature equations that
must be allways satisfied. Thus, from \eqref{q1} we have the relation
\begin{align*}
q_k(s_k+1,s_i-1)-q_i = q_k(s_k+1)-q_i(s_k+1).
\end{align*}
Hence, by \eqref{qrel3} one finds that
\begin{equation*}
 \e_i^{-1}\Psi = \bra{\e_k+q_k(s_k+1,s_i-1)-q_i}^{-1}\Psi =
 \bra{\e_k+q_k(s_k+1)-q_i(s_k+1)}^{-1}\Psi
\end{equation*}
which when use recursively leads to
\begin{multline*}
 \e_i^{-m}\Psi = \bra{\e_k+q_k(s_i+1-m,s_k+1)-q_i(s_i+1-m,s_k+1)}^{-1}\cdot...\\
  ... \cdot \bra{\e_k+q_k(s_k+1)-q_i(s_k+1)}^{-1} \Psi
\end{multline*}
for $m>0$.

\begin{example}The multi-component Toda hierarchy.

Let us use the notation $u(s_k+m):=u^{(m)}$. The linear equation
\eqref{at} represented by $\e_k$ transforms to
\begin{equation*}
\lo \Psi = \bra{\e_k +q_k^{(1)} +a_k\e_k^{-1}+ \sum_{i\neq k}
a_i\bra{\e_k +q_k^{(1)}-q_i^{(1)}}^{-1}} \Psi.
\end{equation*}
Then for $\Psi_{t_{\infty 1}} = (\e_k+q_k^{(1)})\Psi$ one
calculates the multi-component Toda system
\begin{equation*}
\begin{split}
(a_i)_x&= a_i^{(1)}\bra{q_i^{(1)}-q_k^{(1)}} +
a_i\bra{q_k^{(1)}-q_i}\\
(q_i)_x&= \bra{q_i-q_i^{(-1)}}(q_i-q_k) +
\sum_j\bra{a_j-a_j^{(-1)}},
\end{split}
\end{equation*}
i.e. the lattice analogue of \eqref{dtoda2}. For $\Omega_{k 1}\Psi = -a_k\e_k^{-1} \Psi$ we have the counterpart of
\eqref{AKNS2} in the form
\begin{equation*}
\begin{split}
(a_k)_{y_k}&= a_k\bra{q_k^{(1)}-q_k}-a_k\sum_{j\neq
k}\bra{\frac{a_j}{q_j^{(1)}-q_k^{(1)}}-\frac{a_j^{(-1)}}{q_j-q_k}}\\
(q_k)_{y_k}&= a_k-a_k^{(-1)}\\
(a_i)_{y_k}&=
a_k\bra{\frac{a_i}{q_i^{(1)}-q_k^{(1)}}-\frac{a_i^{(-1)}}{q_i-q_k}}\\
(q_i)_{y_k}&= a_k\bra{1-\frac{q_i-q_k}{q_i^{(1)}-q_k^{(1)}}},
\end{split}
\end{equation*}
where $i\neq k$.
\end{example}

\begin{example}\label{exa2}
We will use the notation $q:=q_1$, $\e:=\e_1$, $y:=y_1$ and
$u^{(m)}:=u(s_1+m)$. The linear equation being a dispersive
analogue of \eqref{dbon0} is
\begin{equation}\label{some}
\lo \Psi = \bra{\e + q^{(1)} + u\e^{-1} + v\e^{-2}}\Psi.
\end{equation}
Then, for $\Omega_{\infty 1}\Psi = \bra{\e+q^{(1)}}\Psi$ one finds
\begin{equation*}
 \begin{split}
u_x &= u \bra{q^{(1)}-q}+v^{(1)}-v\\
v_x &=v\bra{q^{(1)}-q^{(-1)}}\\
q_x &= u-u^{(-1)}
 \end{split}
\end{equation*}
analogue of \eqref{aaa}. Next system  is
\begin{align*}
\Omega_{1 1}\Psi = -a\e^{-1}\Psi\qquad \Longrightarrow \qquad
\begin{split}
u_y &= a \bra{q^{(1)}-q}\\
v_y &= u a^{(-1)}-u^{(-1)}a\\
q_y &= a-a^{(-1)}
\end{split}
\end{align*}
where $a$ satisfies $a a^{(-1)}=v$. There are two simplest
possible solutions on the field $a$:
\begin{equation*}
a=\frac{v^{(1)}v^{(3)}v^{(5)}\cdot ...}{v^{(2)}v^{(4)}\cdot
...}\qquad \text{or}\qquad a=\frac{vv^{(-2)}v^{(-4)}\cdot
...}{v^{(-1)}v^{(-3)}\cdot ...}.
\end{equation*}
\end{example}

\begin{example}\label{alg21.4}
In order to complete the dispersionful version of Examples
\ref{alg21.1} or \ref{alg21.2} we need to write the Lax operator
${\lo}$ from Example \ref{alg21.3} in terms of the shift operator
$\e:=\e_1$ as
\begin{align*}
\lo\Psi =
\bra{\e^2+(q^{(1)}+q^{(2)})\e+(q^{(1)})^2+u+a\e^{-1}}\psi.
\end{align*}
Now, from $\Omega_{\infty 1}\Psi = \bra{\e+q^{(1)}}\Psi$ we find
the lattice analogue of \eqref{exq1}
\begin{align*}%\label{exf21}
\bra{q^{(1)}+q^{(2)}}_x &= u^{(1)}-u\\
\bra{(q^{(1)})^2+u}_x &= a^{(1)}-a\\
a_x &= a\bra{q^{(1)}-q}.
\end{align*}
On the other hand, by introducing $b:=q^{(1)}+q^{(2)}$ and
$c:=(q^{(1)})^2+u$, so that $\lo\Psi=\bra{\e^2+b\e+c+a\e^{-1}}\Psi$,
we have also the lattice analogue of \eqref{exq2}
\begin{align*}%\label{exf22}
\Omega_{\infty 2}\Psi = \bra{\e^2+b\e+c}\Psi\qquad \Longrightarrow
\qquad
\begin{split}
b_t &= a^{(2)}-a\\
c_t &= a^{(1)}b-ab^{(-1)}\\
a_t &= a(c-c^{(-1)}).
\end{split}
\end{align*}
\end{example}

As we have been able with the use of expression \eqref{qrel2} to
construct Lax operators $\lo$, being dispersive counterparts of
the meromorphic functions with 'finite' poles of higher order, we
would like to use \eqref{qrel2} in the opposite way and represent
$\lo$ by means of pseudo-differential operators. Thus, one finds the
useful formula
\begin{align}\label{form}
\e_k^{-m} \Psi = (\pr - q_k(s_k-m+1))^{-1}\cdot ... \cdot (\pr -
q_k(s_k-1))^{-1}(\pr - q_k)^{-1}\Psi,
\end{align}
where $m>0$. However, shifted fields will appear, which for
$\alpha=\infty$ can be understood as new independent dynamical
fields. So, one is able to construct closed equations being
finite-field reductions \eqref{algo} of KP hierarchy, also for
$n_i$ higher then $1$, that are analogues of algebraic orbit
reductions of Whitham hierarchy. The above dispersionful analogues
in general will have more dynamical fields then its dispersionless
limits. The reductions of this kind of KP hierarchy have been
considered earlier in \cite{BX1,BX2} in the context of
multi-matrix models.

\begin{example}
The linear equation \eqref{some} written by means of
pseudo-differential operators takes the form
\begin{equation*}
 \lo \Psi = \bra{\pr + u(\pr-q)^{-1} +
 v(\pr-q')^{-1}(\pr-q)^{-1}}\Psi,
\end{equation*}
where $q':=q^{(-1)}$. Then, for $\Omega_{\infty 2}\Psi =
\bra{\pr^2 + 2u}\Psi$ one finds the following system
\begin{equation}\label{some2}
 \begin{split}
u_t &= u_{2x}+2v_x+2(uq)_x\\
v_t &= v_{2x}+2vq_x+2(vq')_x\\
q_t &= -q_{2x}+2u_x+2qq_x\\
q'_t &= -q'_{2x}-2q_{2x}+2u_x+2q'q'_x
 \end{split}.
\end{equation}
The evolutions of fields $q$ and $q'=q(s_1-1)$ in the
quasi-classical limit will be identical, i.e. four-field equation
\eqref{some2} is well defined dispersive counterpart of
three-field \eqref{dbon1}. This system was constructed in
\cite{BX1,BX2}.
\end{example}

In the forthcoming article \cite{S} the theory presented in this
section will be treated more broadly and with all necessary
proofs. The aim of \cite{S} will be a construction of
(1+1)-dimensional field and lattice soliton systems being
dispersive counterparts for a wider class of dispersionless
equations considered in \cite{SB}.

\section{Comments}

In \cite{T} Takasaki provided a scheme for deriving the Whitham
hierarchy as a dispersionless limit of the 'charged'
multi-component KP hierarchy. The charges are introduced through
extra set of discrete variables $s_i$. The scheme is based on the
$\tau$-function bilinear formalism \cite{DJ,KV} where the total
charge $s_\infty + \sum_i s_i$ equal to zero. In this sense the
'charged' multi-component KP hierarchy can be understood as a
multi-component generalization of Toda hierarchy. As a result, the
scalar auxiliary linear equations \eqreff{lin1}{lin2}, defined by
means of shift operators, leading to the Lax formalism
\eqreff{lh}{zerc} can be extracted. In literature there are
several different formulations of the multi-component KP
hierarchy, for example the Grassmannian approach \cite{SW} or the
$\overline{\pr}$-method \cite{BK}. So, it seems that a further
discussion of the corresponding connections between them and the
dispersionful Whitham hierarchy from the present work and
\cite{T,TT1} is needed.

Besides, due to the previous comments on the bi-Hamiltonian
structures of the original and dispersionful Whitham hierarchy we
see that the existence of the symmetries leads to the conclusion
that these hierarchies are incomplete. So, we can extend this
incompleteness onto multi-component KP and Toda hierarchies. This problem seems to be worth of further investigation.

\subsection*{Acknowledgement}

A part of the research was supported by the MISGAM programme of
European Science Foundation (ESF) as well as the MNiI grant no.
N202. The autors are grateful to the Departamento de F\'isica
Te\'orica II of the Universidad Complutense in Madrid for its
hospitality. We are also grateful to Luis Mart\'inez Alonso and
Manuel Ma\~nas for useful discussions on the problem of the
construction of dispesionful Whitham hierarchy and the work
\cite{T} of Kanehisa Takasaki.


\footnotesize
\begin{thebibliography}{99}

\bibitem{K1} I. M. Krichever, {\it The averaging method for two-dimensional "integrable" equations}, Func. Anal. Appl. {\bf 22} (1988) 200-213

\bibitem{K2} I. M. Krichever, {\it The $\tau$-function of the universal Whitham hierarchy,
matrix models and topological field theories}, Comm. Pure Appl.
Math. {\bf 47} (1994) 437-475

\bibitem{GMMA} F. Guil, M. Ma\~nas and L. Mart\'inez Alonso, {\it The Whitham hierarchies:
reductions and hodograph solutions}, J. Phys.A: Math. Gen. {\bf
36} (2003) 4047-4062

\bibitem{K3} I. Kichever, {\it Linear
operator with self-consistent coefficients and rational reductions
of KP hierarchy}, Physica D {\bf 87} (1995) 14-19

\bibitem{EOR} B. Enriquez, A. Yu Orlov and W. N. Rubtsov, {\it Dispersionful analogues of Benney's
equations and $N$-wave systems}, Inverse Problems {\bf 12} (1996)
241-250

\bibitem{OS} W. Oevel and W. Strampp, {\it Constrained KP hierarchy and bi-Hamiltonian structures},
Commun. Math. Phys. {\bf 157} (1993) 51

\bibitem{BX} L. Bonora and C. S. Xiong, {\it Matrix models without scaling limit},
Int. J. Mod. Phys. A {\bf 8} (1993) 2973-2992

\bibitem{BX1} L. Bonora and C. S. Xiong, {\it Multi-field representations of the KP hierarchy
and multi-matrix models}, Phys. Lett. B {\bf 317} (1993) 329

\bibitem{BX2} L. Bonora and C. S. Xiong, {\it The $(N,M)$-th KdV hierrachy and the associated $W$
algebra}, J. Math. Phys. {\bf 35} (1994) 5781-5819

\bibitem{UT} K. Ueno and K. Takasaki, {\it Toda lattice
hierarchy}, Advanced Studies in Pure Math. {\bf 4} (Kinokuniya,
Tokyo, 1984) pp. 1-94

\bibitem{ANP} H. Aratyn, E. Nissimov and S. Pacheva, {\it Constrained KP Hierarchies:
Additional Symmetries, Darboux-B\"acklund Solutions and Relations
to Multi-Matrix Models}, Int. J. Mod. Phys. A {\bf 12} (1997)
1265-1340

\bibitem{CDZ} G. Carlet, B. Dubrovin and Y. Zhang, {\it The
Extended Toda Hierarchy}, Mosc. Math. J. {\bf 4} (2004) 313-332

\bibitem{T} K. Takasaki, {\it Dispersionless integrable hierarchies
revisited}, talk at the
SISSA workshop "Conference on Riemann-Hilbert Problems,
Integrability and Asymptotics" (2005)

\bibitem{TT1} K. Takasaki and T. Takebe, {\it Universal Whitham
hierarchy, dispersionless Hirota equations and multi-component KP
hierarchy}, arXiv:nlin.SI/0608068

\bibitem{STS} M. A. Semenov-Tian-Shansky, {\it What is a classical
r-matrix?}, Funct. Anal. Appl. {\bf 17} (1983) 259

\bibitem{KO} B. G. Konopelchenko and W. Oevel, {\it An r-matrix approach
to nonstandard classes of integrable equations}, Publ. RIMS, Kyoto
Univ. {\bf 29} (1993) 581-666

\bibitem{BM} M. B\l aszak and K. Marciniak, {\it R-matrix approach to
lattice integrable systems}, J. Math. Phys. {\bf 35} (1994) 4661

\bibitem{BS} M. B\l aszak and B. M. Szablikowski, {\it From dispersionless
to soliton systems via Weyl-Moyal-like deformations}, J. Phys A: Math. Gen.
{\bf 36} (2003) 12181-12203

\bibitem{TT} K. Takasaki and T. Takebe, {\it Integrable
hierarchies and dispersionless limit}, Rev. Math. Phys. {\bf 7}
(1995) 743-808

\bibitem{Li} Li Luen-Chau, {\it Classical r-Matrices and Compatible Poisson
Structures for Lax Equations in Poisson Algebras}, Commun. Math.
Phys. {\bf 203} (1999) 573-592

\bibitem{BS2} M. B\l aszak and B. M. Szablikowski, {\it Classical $R$-matrix
theory of dispersionless systems: I. (1+1)-dimension theory}, J. Phys A:
Math. Gen. {\bf 35} (2002) 10325-10344

\bibitem{SB} B. M. Szablikowski and M. B\l aszak, {\it Meromorphic Lax
representations of (1+1)-dimensional multi-Hamiltonian dispersionless systems},
J. Math. Phys.  {\bf 47}  (2006) 92701-92723

\bibitem{MMM} M. Man\~nas, E. Medina and Luis Mart\'inez Alonso,
{\it On the Whitham hierarchy: dressing scheme, string equations
and additional symmetries},  J. Phys. A: Math. Gen. {\bf 39}  (2006) 2349-2381

\bibitem{EY} T. Eguchi and S-K Yang, {\it The Topological $CP^1$ Model and
the Large-$N$ Matrix Integral}, Mod. Phys. Lett. A {\bf 9} (1994) 2893-2902

\bibitem{AK} S. Aoyama and Y. Kodama, {\it Topological Landau-Ginzburg Theory
with a Rational Potential and the Dispersionless KP Hierarchy},
Commun. Math. Phys. {\bf 182} (1996) 185-219

\bibitem{S} B. M. Szablikowski, {\it Meromorphic-like Lax
representations of (1+1)-dimensional field and latices soliton systems}, preprint

\bibitem{DJ} M. Jimbo and T. Miwa, {\it Solitons and infinite dimensional Lie
algebras}, Publ. Res. Ins. Math. Sci. {\bf 19} (1983) 943-1001

\bibitem{KV} V. G. Kac and J. W. van de Leur, {\it The $n$-component KP hierarchy and representation theory}, J. Math. Phys. {\bf
44} (2003) 3245-3293


\bibitem{SW} G. Segal and G. Wilson, {\it Loop groups and equations
of KdV type} Inst. Hautes Etudes Sci. Publ. Math {\bf 63} (1995)
1-64

\bibitem{BK} L. V. Bogdanov  and B. G. Konopelchenko, {\it Analytic-bilinear approach to integrable hierarchies. II. Multicomponent KP and 2D Toda lattice hierarchies}, J. Math. Phys. {\bf 39} (1998) 4701-4728


\end{thebibliography}
\end{document}